\documentclass[10pt,conference]{IEEEtran}
\usepackage[utf8]{inputenc}

\usepackage{balance}
\usepackage{booktabs}
\usepackage{multirow}
\usepackage[noadjust]{cite}
\usepackage{hyperref} 
\usepackage{soul}
\usepackage{xcolor}
\usepackage{xspace}
\usepackage{url}
\usepackage{graphicx}
\usepackage{listings,multicol}
\usepackage[caption=false]{subfig}  
\usepackage[export]{adjustbox}
\usepackage{amsmath}
\usepackage{amssymb}
\usepackage{textcomp}
\usepackage{breakurl} 
\usepackage{listings}
\usepackage{cleveref}
\usepackage{algorithm}
\usepackage[noend]{algpseudocode}
\algnewcommand\algorithmicinput{\textbf{Input:}}
\algnewcommand\algorithmicoutput{\textbf{Output:}}
\algnewcommand\Input{\item[\algorithmicinput]}%
\algnewcommand\Output{\item[\algorithmicoutput]}%

\newcommand{\toolname}{\textsc{\textbf{DapStep}}: \textbf{D}eep \textbf{A}ssignee \textbf{P}rediction \\ for \textbf{S}tack \textbf{T}race \textbf{E}rror re\textbf{P}resentation}

\newcommand{\secpart}[1]{\subsection{#1}}
\newcommand{\subsecpart}{\subsubsection}

\makeatletter
\newcommand{\linebreakand}{%
  \end{@IEEEauthorhalign}
  \hfill\mbox{}\par
  \mbox{}\hfill\begin{@IEEEauthorhalign}
}
\makeatother

\newcounter{observation}
\newcommand{\observation}[1]{\refstepcounter{observation}
	\begin{center}
		\framebox{
			\begin{minipage}{0.93\columnwidth}
				{\bf Answer to RQ\arabic{observation}:} \textit{#1}
			\end{minipage}
		}
	\end{center}
}

\title{\toolname}

\author{
\IEEEauthorblockN{Denis Sushentsev}
\IEEEauthorblockA{
    \textit{HSE University}\\
    \textit{JetBrains}\\
    Saint Petersburg, Russia \\
    denis.sushentsev@jetbrains.com
}
\and
\IEEEauthorblockN{Aleksandr Khvorov}
\IEEEauthorblockA{
    \textit{HSE University}\\
    \textit{JetBrains}\\
    Saint Petersburg, Russia \\
    aleksandr.khvorov@jetbrains.com
}
\and
\IEEEauthorblockN{Roman Vasiliev}
\IEEEauthorblockA{
    \textit{JetBrains}\\
    Saint Petersburg, Russia \\
    roman.vasiliev@jetbrains.com
}
\linebreakand
\IEEEauthorblockN{Yaroslav Golubev}
\IEEEauthorblockA{
    \textit{JetBrains Research}\\
    Saint Petersburg, Russia \\
    yaroslav.golubev@jetbrains.com
}
\and
\IEEEauthorblockN{Timofey Bryksin}
\IEEEauthorblockA{
    \textit{JetBrains Research}\\
    \textit{HSE University}\\
    Saint Petersburg, Russia \\
    timofey.bryksin@jetbrains.com
}
}

\begin{document}

\maketitle

\begin{abstract}
    The task of finding the best developer to fix a bug is called \textit{bug triage}. Most of the existing approaches consider the bug triage task as a classification problem, however, classification is not appropriate when the sets of classes change over time (as developers often do in a project). Furthermore, to the best of our knowledge, all the existing models use textual sources of information, \textit{i.e.,} bug descriptions, which are not always available.
    
    In this work, we explore the applicability of existing solutions for the bug triage problem when stack traces are used as the main data source of bug reports. Additionally, we reformulate this task as a ranking problem and propose new deep learning models to solve it. The models are based on a bidirectional recurrent neural network with attention and on a convolutional neural network, with the weights of the models optimized using a ranking loss function. To improve the quality of ranking, we propose using additional information from version control system annotations. Two approaches are proposed for extracting features from annotations: manual and using an additional neural network. To evaluate our models, we collected two datasets of real-world stack traces. Our experiments show that the proposed models outperform existing models adapted to handle stack traces. To facilitate further research in this area, we publish the source code of our models and one of the collected datasets.
\end{abstract}

\section{Introduction}\label{sec:introduction}

Software bugs are an inevitable part of the development process. Bugs can lead to security problems, loss of company profit, and in the worst case, even fatal accidents~\cite{wong2017more}. For these reasons, bugs need to be swiftly fixed, which requires choosing the most appropriate developer. The problem of finding such a developer for a particular bug is called \textit{bug triage}~\cite{Anvik2006WhoSF}.

The developer who should fix the bug can be assigned manually, however, such an approach has several significant disadvantages. Firstly, it is tedious and time-consuming work, and the situation gets more and more complicated as the number of developers grows. In large companies, hundreds of bug reports are received every day, which makes manual developer assignment very difficult if not impossible. For example, 333,371 bugs were reported for the Eclipse IDE from October 2001 to December 2010, averaging at about 100 bugs every day~\cite{Xuan2012DeveloperPI}. Secondly, it is important to assign the most suitable developer right from the start to reduce the time of bug fixing~\cite{Anvik2006WhoSF}. Otherwise, the error gets reassigned from one developer to another~\cite{Jeong2009ImprovingBT}, and as a result, the time of each developer in such a chain is wasted, while the error remains in the system longer, which can be critical. 

A large number of approaches have been proposed to solve the bug triage problem automatically. Existing models can be roughly divided into three groups: based on heuristics~\cite{Tian2016LearningTR, Tamrawi2011FuzzySA, Shokripour2015ATA, Hu2014EffectiveBT}, based on classic machine learning~\cite{Naguib2013BugRA, Xia2017ImprovingAB, sarkar2019improving}, and based on deep learning (DL)~\cite{Lee2017ApplyingDL, Guo2020DeveloperAM, Mani2019DeepTriageET, Xi2019BugTB}. The works of Guo et al.~\cite{Guo2020DeveloperAM} and Mani et al.~\cite{Mani2019DeepTriageET} demonstrated that deep learning helps with the task of assigning a developer better than other approaches.  This is to be expected, since the bug triage task is based on natural language processing, where deep learning shows promising results~\cite{Wolf2020TransformersSN}. An additional advantage of deep learning algorithms is that they do not require sophisticated feature extraction methods~\cite{Schmidhuber2015DeepLI}.

\begin{figure}[h]
    \centering
    \includegraphics[width=\columnwidth]{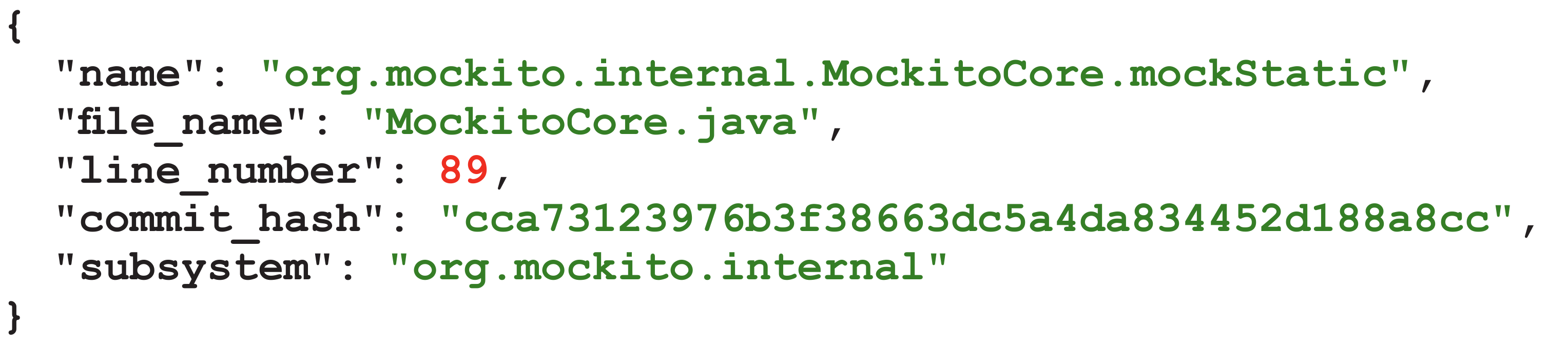}
    \vspace{-0.5cm}
    \caption{An example of a stack frame. The frame consists of the name of the function that led to the error, as well as various information about it.}
    \label{fig:introduction-frame-example}
\end{figure}

However, it should be noted that bugs can be reported in different forms. For example, in a bug tracking system, errors are usually present in the form of a bug report: a name, a small description in some natural language, and some additional meta information (the date the error was introduced, priority, severity, etc.). To the best of our knowledge, all the existing solutions are based on working with this kind of error representation. At the same time, errors can also come in the form of \textit{stack traces}: sequences of function calls (called \textit{frames}) that lead to an error in the system. Developers commonly use stack traces during debugging, and users can usually see a stack trace displayed as part of an error message. Stack traces help to solve the bug localization problem~\cite{Wong2014BoostingBF, Moreno2014OnTU, Youm2015BugLB} and the bug report deduplication problem~\cite{Dang2012ReBucketAM, Sabor2017DURFEXAF, Khvorov2021S3MSS}. The example of a single stack frame is presented in \Cref{fig:introduction-frame-example}. 

Stack traces are a data source that is often easy to obtain: most modern software systems are able to automatically send back stack traces of the error that has occurred. In such a setting, predicting the assignee by the textual description of the error would require labeling all the error reports, which is almost impossible since the number of such reports per day could be enormous. Another important reason to process stack traces automatically is that they are more complicated to analyze manually by people who did not participate in the development of a particular system component, since the information is presented in a rather raw form. Thus, a new approach is needed that solves the bug triage problem for the case where only the stack trace information is available.

Another important limitation of the existing approaches is that they consider the bug triage task as a classification problem. The classification setting might not be the best choice in practice, since the set of classes (developers) can change over time: developers can leave and join the team responsible for the product of even the company itself. 

To the best of our knowledge, no one has previously suggested using bug stack traces as the main source of information for the bug triage problem. In this work, we strive to fill this gap in research to support working with the systems where stack traces are the primary type of data. To that end, we collected two datasets of real-world bug stack traces from JetBrains,\footnote{JetBrains: \url{https://www.jetbrains.com/}} the developer of a wide array of software products including IntelliJ-based IDEs. The larger dataset contains 11,139 stack traces, however, it contains proprietary company code, so we also curate the second dataset --- a smaller public subset of the first one that contains 3,361 stack traces that we release for researchers and practitioners. The datasets consist of a labeled set of bug reports and annotations from the version control system (developer IDs and timestamps) that we apply to improve the quality of our model.

We propose a new approach to solve the bug assignee prediction problem based on stack traces --- a DL-based ranking model called \textit{DapStep} (an RNN ranking model with manual frame-based \& stack-based features). We compared the proposed model with existing approaches adapted for stack trace processing. The proposed model shows Acc@1 of 0.34 and MRR of 0.43 on the public dataset and Acc@1 of 0.60 and MRR of 0.70 on the private dataset.

The main contributions of this paper are as follows:
\begin{itemize}
    \item We propose bug stack traces as a self-sufficient source of information for the assignee prediction task and carry out the first study in comparing various approaches in this setting.
    \item We introduce two bug triage ranking models based on recurrent neural networks (RNN) with the attention mechanism and convolutional neural networks (CNN). The models outperform the existing classification approaches by
    15--20 percentage points on the public dataset, and 17--18 percentage points on the private dataset.
    \item We publish the source code of all the studied models, as well as the public dataset, for future researchers and practitioners: \url{https://github.com/Sushentsev/DapStep}.
\end{itemize}

The remaining sections of this paper are organized as follows. \Cref{sec:related-work} provides a brief overview of existing solutions, and in \Cref{sec:approach}, we propose a new deep learning solution. We evaluate our approach in \Cref{sec:evaluation}, followed by a discussion of the threats to validity in \Cref{sec:threats-to-validity}. Finally, \Cref{sec:conclusion} summarizes the results of the paper.
\section{Related work}\label{sec:related-work}

The bug triage task is a well-established area of research, with a large number of proposed approaches. Previous works can be broadly divided into three large groups: based on \textit{heuristics}, on \textit{classic machine learning}, and on \textit{deep learning}.

Heuristic-based approaches tend to consider the relevance scores of developers and errors based on domain knowledge. Kagdi et al.~\cite{Kagdi2012AssigningCR}, Shokripour et al.~\cite{Shokripour2012AutomaticBA, Shokripour2013WhySC}, and V{\'a}squez et al.~\cite{Vsquez2012TriagingIC} use the information about code authorship, commit messages, comments in the source code, etc. Also, various indexing and NLP techniques are used to search for files related to the query bug report. The most appropriate developers are then selected based on their activities in the relevant files.

Since the software development process is impossible without team work, developers often interact with each other. The result is a collaboration network that can be used as another source of information. Hu et al.~\cite{Hu2014EffectiveBT} and Zhang et al.~\cite{Zhang2013AHB} use collaboration networks and information retrieval techniques on graphs to choose the most appropriate developer. 

As the influence of machine learning spread, it became actively applied in the assignee recommendation as well. Often, such approaches vectorize the text of the bug summary and description using TF-IDF or Bag-of-words (BOW), and classify them using a machine learning algorithm: Naive Bayes, Random Forest, or SVM~\cite{Anvik2006WhoSF, Lin2009AnES, Banitaan2013TRAMAA, Ahsan2009AutomaticSB}. 

Recently, deep learning solutions also became popular. 
Lee et al.~\cite{Lee2017ApplyingDL} present one of the first DL models based on the CNN and Word2Vec embeddings used for assigning a developer to fix the bug. Their approach achieved higher accuracy in industrial projects at LG compared to an open source project.

The application of CNN for the bug triage problem has been reported to be useful in more recent approaches. Guo et al.~\cite{Guo2020DeveloperAM} compare the CNN-based model to the models based on Naive Bayes, SVM, kNN, and Random Forest. The experimental results show that the CNN-based approach outperforms other solutions. Since some of the developers can change jobs or leave the company indefinitely, the authors also propose to reorder developers based on their activity.

Zaidi et al.~\cite{Zaidi2020ApplyingCN} explore different word embeddings for the CNN model: Word2Vec~\cite{Mikolov2013EfficientEO}, GloVe~\cite{Pennington2014GloVeGV}, and ELMo~\cite{Peters2018DeepCW}. The experimental results suggest that the ELMo embeddings are the best for the bug triage problem. 

Chen et al.~\cite{Chen2019AnEI} extend the work on incident triaging (unplanned interruptions or outages of the service) and perform an empirical study on the datasets provided by Microsoft. They explore different bug triage techniques: based on machine learning, deep learning, topic modeling, tossing graphs, and fuzzy sets. On average, the DL technique performs best. 

An alternative to CNNs are RNNs, which are one of the most popular and effective approaches for processing sequences of variable length. Mani et al.~\cite{Mani2019DeepTriageET} use RNNs for assigning the developer to fix a bug. To address the common issue of RNNs ``forgetting'' long sequences~\cite{Hochreiter2001GradientFI}, they propose to apply a bidirectional network with an attention mechanism. Moreover, the neural network learns syntactic and semantic features in an unsupervised manner, which means that it has the ability to use unfixed bug reports. Their work shows that the proposed approach provides a higher average accuracy rank than BOW features with softmax classifier, SVM, Naive Bayes, and cosine distance.

Finally, Xi et al.~\cite{Xi2019BugTB} propose to use a bug tossing sequence to improve the DL model that helps to reassign the bug if the assignment was incorrect. The proposed approach was evaluated on three different open-source projects and outperforms baseline RNN and CNN models. 

In our work, we strive to overcome the limitations of the existing approaches: namely, their reliance on textual descriptions and their use of classification models.
\section{Approach}\label{sec:approach}

In this section, we describe our algorithm for assignee prediction. We consider bug triage as a ranking problem, which we believe to be more appropriate here, because it does not depend on the current set of developers. The classifying setting requires an immutable set of predicted classes. If a developer leaves the project, they should be filtered out of the resulting prediction afterwards, and when a new developer joins, the classifying model will have to be retrained to take them into account. Since developers in the project may come and go, a more suitable option is the ranking problem setting, in which it is necessary to evaluate the relevance function $f(q, d)$ for a bug $q$ and a developer $d$.

More formally, given a query $q$ (bug) and a collection $D$ of documents (developers) that match the query, the task is to find a function $f$ such that $(q, d) \prec (q, d') \Leftrightarrow f(q, d) < f(q, d')$, where $(q, d) \prec (q, d')$ means that $d$ has a rank lower than $d'$. Function $f$ maps query-documents pairs to a relevance score.

The proposed model uses bug stack traces as the primary source of information for predicting assignees. In order to obtain better results, we also build features from the version control system (VCS) annotations, which provide information on which developer modified each line of the file and when. For example, Git annotations can be obtained via the \textit{git blame} or \textit{git annotate} commands.

\begin{figure}[t]
    \centering
    \includegraphics[width=0.7\columnwidth]{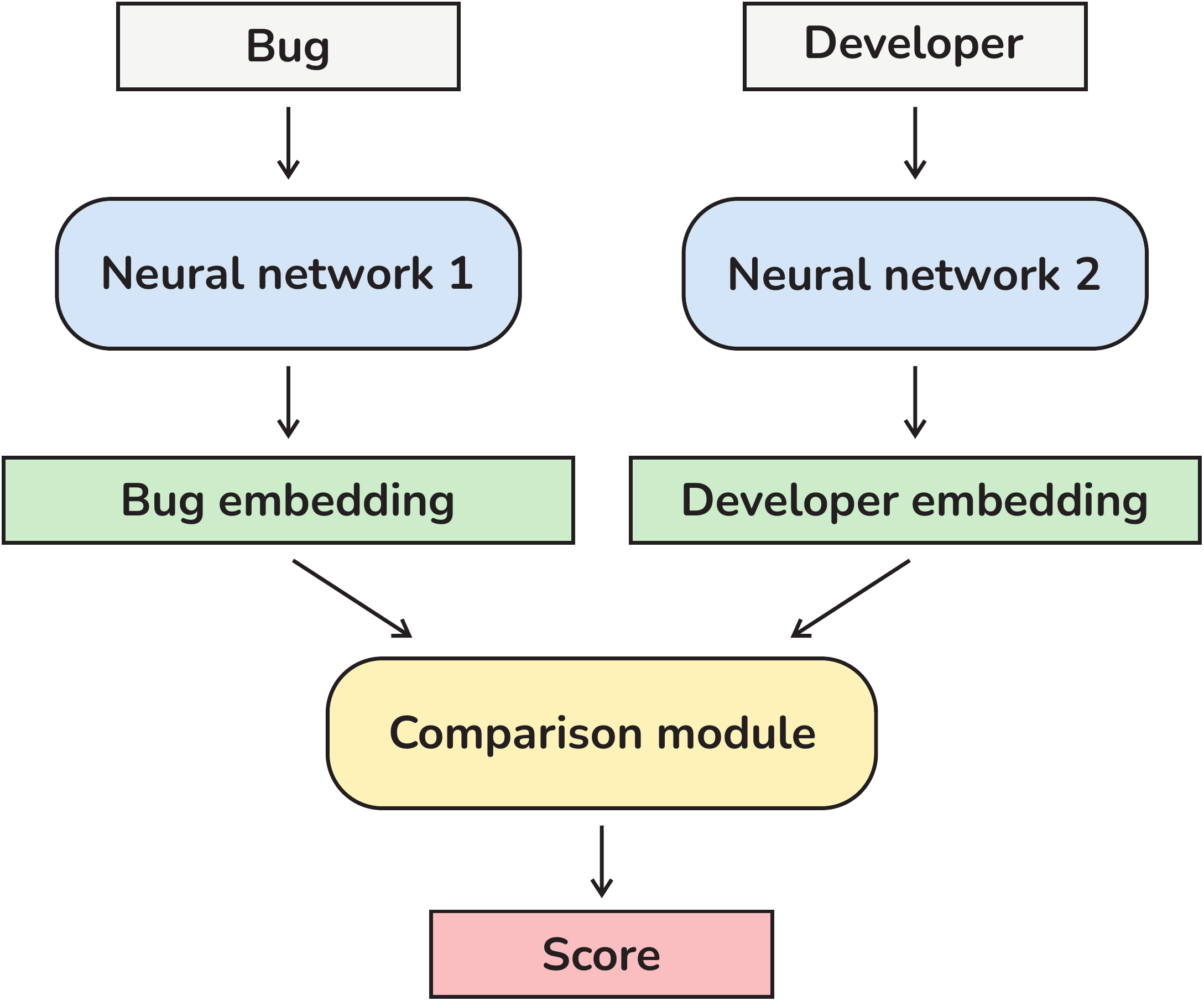}
    \centering
    \caption{The overall pipeline of the approach.}
    \label{fig:approach-architecture}
\end{figure}

The overall pipeline of the proposed algorithm is presented in \Cref{fig:approach-architecture}. Using deep learning methods, the bug and the developer are mapped to a vector of a fixed size (embedding). We transform each bug stack trace into a sequence of text tokens (\Cref{sec:preprocessing}) and apply the ideas from text sequences processing to obtain embeddings of bugs (\Cref{sec:vector-representation}). Then, to create an embedding of a developer, we process all the files in the given stack trace to find files that the developer edited, and use this information to map this developer into the stack trace embedding space (\cref{sec:dev-representation}). After all the embeddings are extracted, they are compared using the comparison module (\Cref{sec:vector-similarity}), and the score is obtained, which shows the relevance of the bug and the developer. To get the most appropriate developers for a given bug, we simply have to rank all the developers by their score. 

To improve our model, we use additional features based on the VCS annotations and propose to process the annotations in two different ways: manually (\Cref{sec:manual-features}) and using an additional neural network (\Cref{sec:neural-features}) that allows us to avoid complex feature engineering. 
Let us now describe these steps in greater detail.

\secpart{Preprocessing}
\label{sec:preprocessing}

The stack trace is represented as a sequence of frames $ST = \{ f_1, f_2, \ldots, f_n \}$, where $f_i$ is the $i$-th stack frame. Every frame has a method name, a file name, a subsystem name, a commit hash, and an error line. An example of a stack frame is presented in~\Cref{fig:introduction-frame-example}.

Our preliminary experiments showed that stack trace preprocessing is an essential step that can significantly improve the model quality. In our work, we used the following data processing steps.

Firstly, we noticed that the length of the stack trace can sometimes be quite large. For instance, the maximum stack trace length in our dataset reached as many as 15,000 frames. It is difficult to make a neural network remember all the information as the frames are processed one by one. On the other hand, long stack traces tend to relate to a \textit{StackOverflowException} error. Oftentimes, such a stack trace contains a loop: a set of frames that repeat at a specific frequency. Replacing the loop with the first occurrence of the loop element allows us to significantly reduce the length of the trace stack without degrading the model's quality. We did this for every stack trace in the dataset where it is applicable.

Secondly, because of the way the dataset was collected, not all information is available for every frame, the frame fields can be null. If the text token received from the frame is null, then we skip this frame.

In order to apply existing approaches, we propose to represent a stack trace as a sequence of text tokens using the following technique. Firstly, we extract the method name, the file name, or the subsystem name from each frame of the stack trace. For example, the stack frame in \Cref{fig:introduction-frame-example} can be mapped to \textit{org.mockito.internal.MockitoCore.mockStatic}, \textit{MockitoCore.java}, or \textit{org.mockito.internal}, respectively. Thus, the stack trace will be presented as a sequence of text tokens, which can be processed with various deep learning approaches. We conducted experiments with all three options (method name, file name, or subsystem), and since the difference was insignificant, we decided to extract the stack trace file name.

\secpart{Representing Stack Traces as Vectors}
\label{sec:vector-representation}

To represent a stack trace with a vector of a fixed length (\textit{i.e.}, embedding), we were inspired by the architectures applied in the previous works, namely, RNNs with attention and CNNs~\cite{Lee2017ApplyingDL, Guo2020DeveloperAM, Zaidi2020ApplyingCN, Mani2019DeepTriageET}. These two types of neural networks are among the most popular in the natural language processing field. In our study, we experimented with both of them.

\subsecpart{Recurrent Neural Network}

An RNN architecture called LSTM~\cite{Hochreiter1997LongSM} is frequently used to handle sequential data. It takes a sequence of text tokens as input and produces the resulting vectors. However, LSTMs may have problems remembering long sequences~\cite{Vaswani2017AttentionIA}, which can be fixed with a bidirectional network~\cite{Graves2005BidirectionalLN} with attention. The attention technique allows to focus on important parts of the input data~\cite{Bahdanau2015NeuralMT}. For instance, frames that are at the top of the stack trace are usually more informative and useful. 

We use the neural network architecture from the work of Maini et al.~\cite{Mani2019DeepTriageET}. The input of the model is a sequence of vector representations of words, $\mathbf{x} = \{ \mathbf{x_1}, \mathbf{x_2}, \ldots, \mathbf{x_n} \}$. In our approach, we use trainable embeddings for every text token. The network is bidirectional, therefore, the sequence is processed in both directions. The RNN produces a sequence of outputs $\mathbf{y} = \{\mathbf{y_1}, \mathbf{y_2}, \ldots, \mathbf{y_n} \}$ from each direction. After that, the attention mechanism is applied, which is the weighted sum of the RNN outputs:
\begin{align}
    \mathbf{a_n} = \sum_{i = 1}^{n} \alpha_i \mathbf{y_i},
\end{align}
where $\alpha_i$ represents an attention weight for the $i$-th output vector.

The final representation $\mathbf{r}$ is obtained as follows:
\begin{align}
    \mathbf{r} = \underbrace{\mathbf{y_n} \oplus \mathbf{a_n}}_\text{forward LSTM} \oplus \underbrace{\mathbf{y_n} \oplus \mathbf{a_n}}_\text{backward LSTM},
\end{align}
where $\oplus$ represents the concatenation of vectors. It is easy to see that if the output vector has dimension $d$, then the embedding $r$ will be of size $4 \times d$.

\subsecpart{Convolutional Neural Network}

Another possible approach to represent a stack trace with a vector is to use CNN. CNNs are most commonly applied to analyze visual information, however, they can also solve natural language processing tasks~\cite{Collobert2008AUA}.

In a CNN-based network, for each sequence of text tokens, we build a matrix $\mathbf{S} \in \mathbb{R}^{s \times d}$, where $s$ is the sequence length and $d$ is the embedding dimension. We were inspired by the work of Lee et al.~\cite{Lee2017ApplyingDL} when building the model architecture. Similarly to them, we use trainable embeddings for text tokens. After that, a convolution layer with 1D convolutions is used to extract different patterns from the sequence of tokens. After applying each convolution filter, a feature vector is obtained. In the extracted feature vector, the subsampling process called max-pooling is applied, which is the operation of extracting the maximum element from a vector. The final representation $\mathbf{r}$ is obtained by concatenating max-pooling values and has a dimension equal to the total number of convolutions.

\secpart{Representing Developers as Vectors}
\label{sec:dev-representation}
Obtaining an embedding of a given bug is pretty straightforward, since each bug has a stack trace that can be transformed into a sequence of text tokens. However, the process of extracting the embedding of a developer is not that obvious. 

One possible solution is to represent the developer as all the code they wrote in the system. This approach has a significant drawback: the need to regularly re-index a large amount of data. If the developer has written new code in the system, then this must be taken into account. Continuous and efficient updates of the developer's embedding is a challenging task. 

To address this problem, we propose to map every developer to a specific synthetic stack trace, more specifically, a sequence of stack frames that they edited. In order not to deal with large-scale re-indexing, we do not use all the available stack traces, but only the stack trace of the current (query) bug. This way, the developer embedding will be bug-dependent: different vector representations are built for different errors, there is no single developer representation. This approach allows us to build the embedding of a developer much faster. The average length of a stack trace in our datasets is 50 frames, therefore, it is enough to look at about 50 files in order to map the developer to their stack trace. Furthermore, the resulting ``developer stack trace'' can be handled in the exact same way as the bug stack trace, and it is possible to use the same network architecture for the bug and for the developer, because each of them is represented in the same form. 

\Cref{alg:developer-stacktrace} shows the pseudo-code for the building of this developer stack trace. In this algorithm, we look at all frames from the stack trace of the current bug from first to last. If the developer has edited at least one line from the file of the given frame, then this frame is included in the developer stack trace. Each stack trace is an ordered sequence of frames, they are numbered starting from the top of the stack. While building the developer stack trace, the order of the frames is preserved. The order of the frames is significant, because generally frames at the top of the stack are more revealing.

\begin{algorithm}
    \caption{The building of the developer stack trace.}\label{alg:developer-stacktrace}
        \begin{algorithmic}
            \Input Developer $dev$, stack trace $stack$
            \Output Developer stack trace $dev\_stack$
            \State dev\_stack $\gets$ emptyList
            
            \For{frame in stack}
                \State file $\gets$ getFrameFile(frame)
                \State authors $\gets$ getFileAuthors(file)
                
                \If{dev $\in$ authors}
                    dev\_stack.append(frame) 
                \EndIf
                
            \EndFor
            \Return dev\_stack
        \end{algorithmic}
\end{algorithm}

It is important to note that the inner frames in the stack trace can include files from various libraries, in which case they will not have been edited by any of the developers in the project. We leave dealing with this case specifically for future work, for example, it might be possible to use the history of the developer's work to see whether they fixed bugs that relate to this particular library.

Overall, for each bug and each developer, we obtain a special stack trace that contains only the frames that concern files that this developer has edited. This allows us to compare the resulting embeddings.

\begin{table*}
    \centering
    \caption{Additional features obtained from the VCS annotations and their normalizations.}
    \label{table:annot-features}
    \begin{tabular}{cll}
        \toprule
        \multicolumn{1}{c}{\textbf{Category}} & 
        \multicolumn{1}{c}{\textbf{Description}} & \multicolumn{1}{c}{\textbf{Normalizations}} \\ 
        \midrule
        \multirow{8}{4em}{\textbf{Frame}} 
        & 
        Minimum distance from the edited line to the error line & Raw; annotation length; min \\
        & 
        Did the developer edit the error line? & Raw  \\
        & 
        Normalized number of edited lines in the file & Annotation length; max \\
        & 
        Normalized number of edited lines weighted by time & Annotation length; max \\ 
        & 
        Normalized number of edited lines in the window of size 10 & Window size; max \\
        & 
        Number of different developer's timestamps & Raw; max \\
        &
        Time passed since the last edit & Exp(-x); Log(x) \\ 
        & 
        Time passed since the first edit & Log(x) \\
        \midrule
        \multirow{5}{4em}{\textbf{Stack}} 
        & 
        The order of the first edited frame & Raw; stack length; number of annotated frames; min \\
        & 
        Normalized number of edited error lines & Stack length; max \\
        & 
        Normalized number of edited lines & Total number of lines; max \\
        &  
        Normalized number of edited lines in the frame with maximum IDF & Annotation length \\
        & 
        Normalized number of edited frames & Stack length; number of annotated frames; max \\
        \bottomrule
    \end{tabular}
\end{table*}

\secpart{Additional Features}
\label{sec:manual-features}

To improve the performance of the model, we enrich the embedding with the features built from the VCS annotations. 
The annotations provide information about who was the last person to have changed each line in the file, and when this change took place. The rationale behind using annotations is the following: if a developer has recently edited some file, it is more likely that their changes resulted in a bug. Therefore, such a developer should probably fix the current bug. 

\begin{figure}[htbp]
    \centering
    \includegraphics[width=\columnwidth]{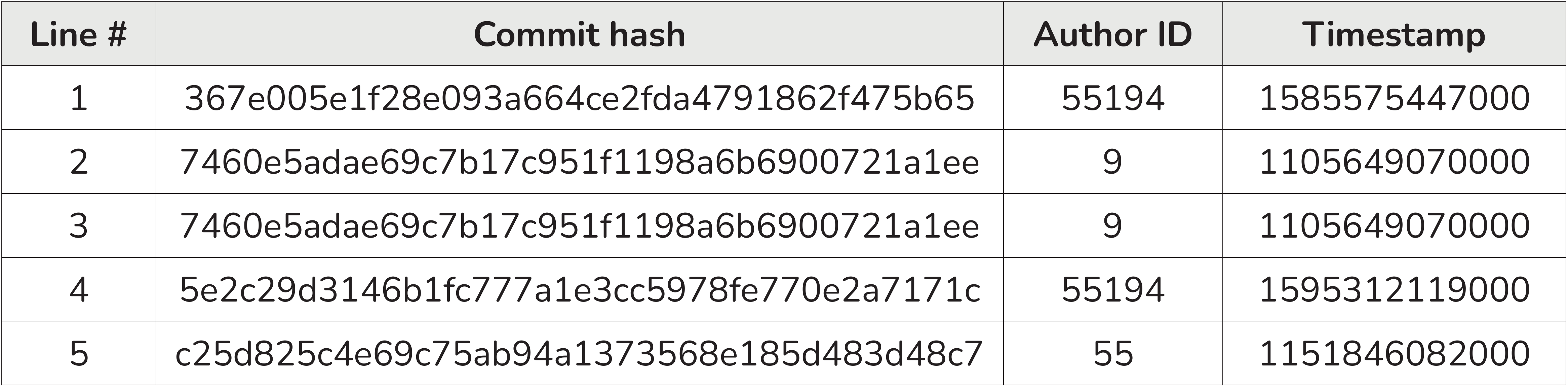}
    \centering
    \vspace{-0.2cm}
    \caption{An example of the first lines of an annotation.}
    \label{fig:annotation-example}
\end{figure}

An example of the first five lines of an annotation is shown in \Cref{fig:annotation-example}. Each developer is encoded with a unique identifier, and the time is represented in the Unix epoch format. 

Additional features can be constructed both on the level of individual stack frame (\textit{e.g.}, how many lines in the file of a specific frame the developer edited) and on the level of the entire stack trace (\textit{e.g.}, how many stack frames have files that the developer edited), and are applied in different ways.

\begin{figure}[htbp]
        \centering
        \includegraphics[width=\columnwidth]{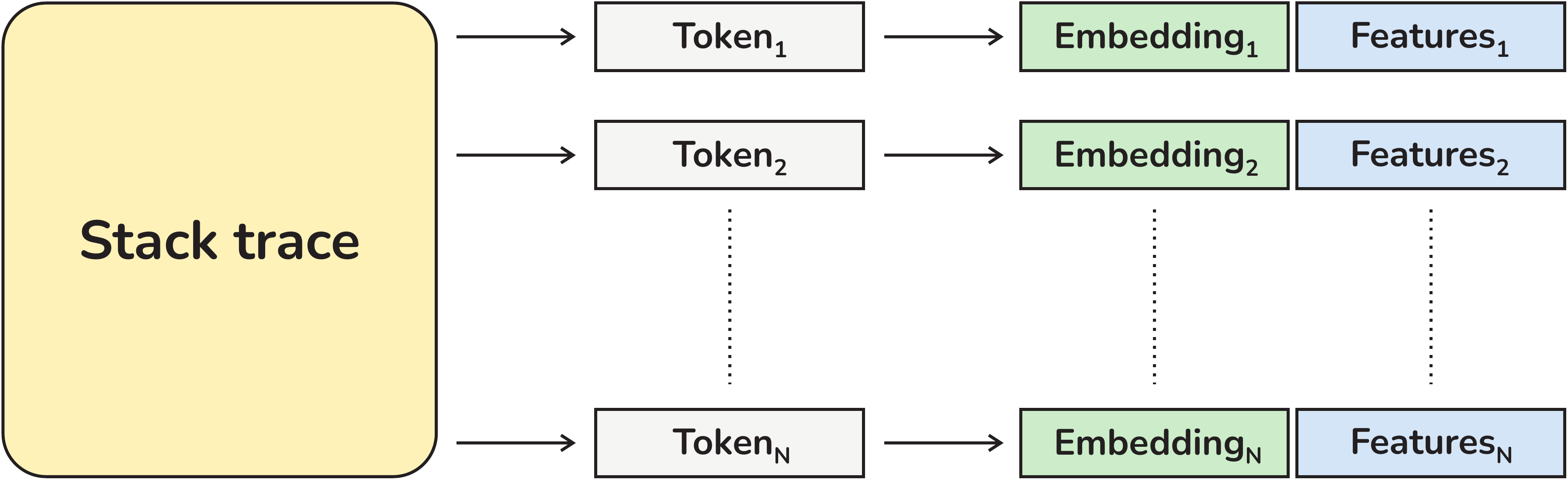}
        \vspace{-0.4cm}
        \centering
        \caption{The application of the frame-based features.}
        \label{fig:approach-frame-features}
    \end{figure}

Features that relate to individual frames can be concatenated to the trainable embeddings before applying the RNN (\Cref{sec:vector-representation}). \Cref{fig:approach-frame-features} shows the proposed approach: a text token is extracted from the frame, each text token is associated with a trainable embedding, and the additional feature vector is concatenated to the embedding. The resulting vector becomes the input of the RNN.
    
\begin{figure}[htbp]
        \centering
        \includegraphics[width=\columnwidth]{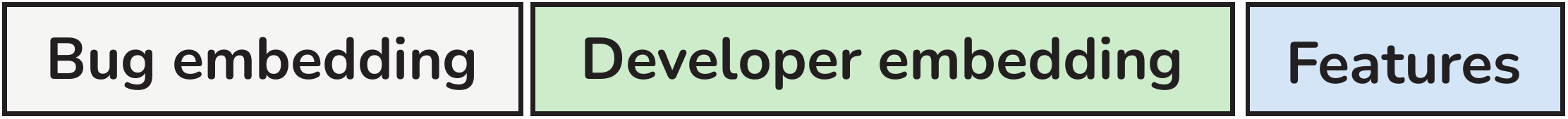}
        \vspace{-0.4cm}
        \centering
        \caption{The application of the stack-based features.}
        \label{fig:approach-overall-features}
    \end{figure}

The features that relate to the entire stack trace can be concatenated to the bug embedding and the developer embedding as presented in \Cref{fig:approach-overall-features}. The resulting vector is the input of the comparison module. 
    
We performed feature engineering on the private dataset, trying different combinations of metrics and their normalization methods. We ended up with 15 frame features and 24 stack trace features that worked best in our setting. They are presented in \Cref{table:annot-features}. 
For example, from the first line of the table, we get three different features: a raw value of the minimum distance and two normalizations (by annotation length and by the minimum value).

\secpart{Neural Annotation Processing}
\label{sec:neural-features}

Manual feature engineering is a complex process that requires domain knowledge and expertise. As an alternative, we also propose using another neural network to extract features from annotations automatically.

The idea behind the annotation processing is as follows: each line of an annotation is labelled with a timestamp of its last change. We suggest to represent annotation lines as elements of a time series --- a sequence of values indexed in the chronological order. We propose to use the distance from the current line to the error line (simply subtracting the line numbers) as the values of the time series, and timestamps of the last modification as the corresponding time.
The considered time series is irregular: code lines could be changed at any time. 
Since this is the first work using DL-based annotation processing, we decided to start with simple things first and use the most popular and straightforward solution for irregular time series processing: concatenate the time information to the time series value to form a vector of size 2.

\begin{figure}[t]
    \centering
    \includegraphics[width=\columnwidth]{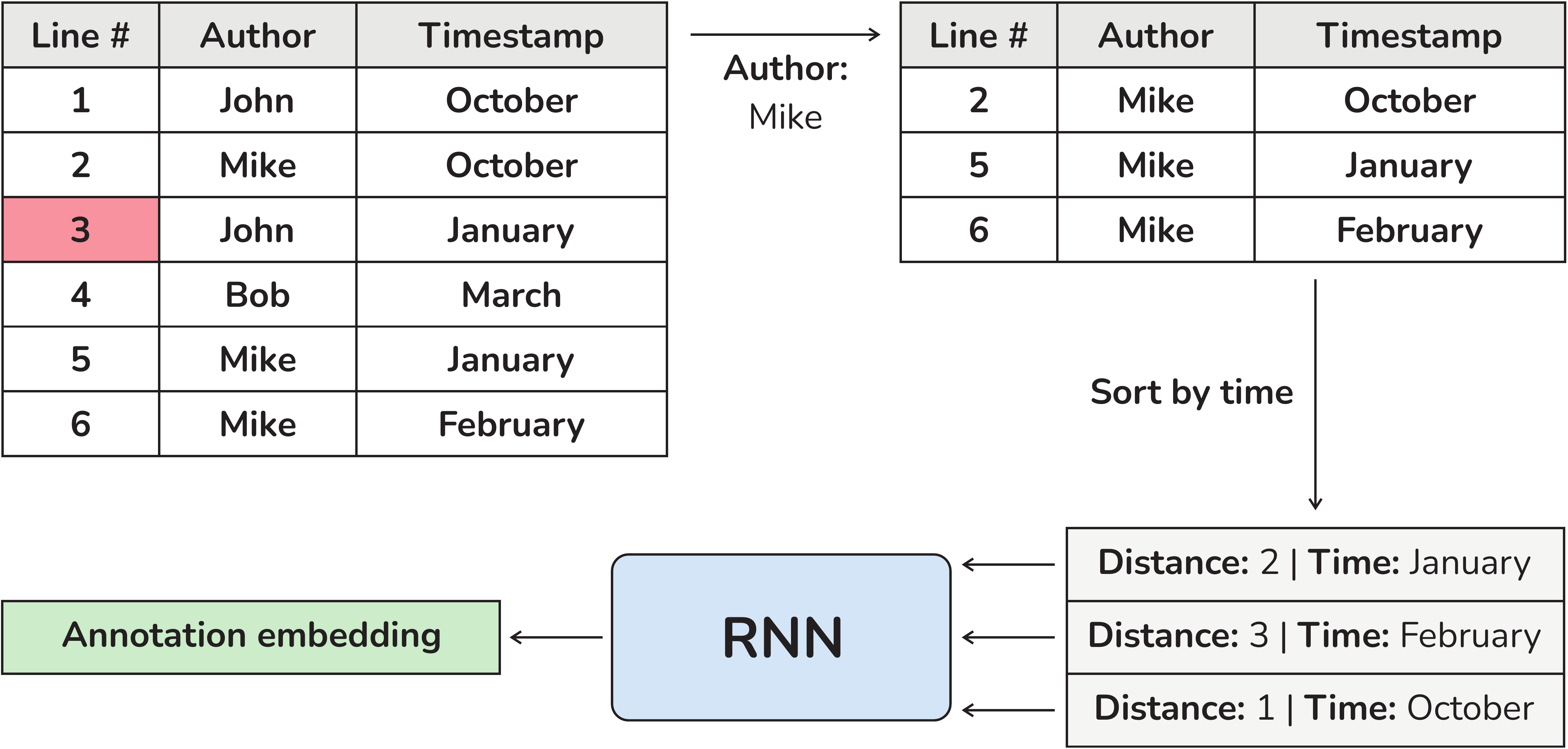}
    \centering
    \caption{The example processing of an annotation.}
    \label{fig:approach-annotation-embedding}
\end{figure}

\Cref{fig:approach-annotation-embedding} shows an example of the annotation processing for developer \textit{Mike}, this will be done for each developer and for each stack frame: 

\begin{itemize}
    \item Select lines from the annotation that were edited by Mike.
    \item Sort the lines by time. Each annotation line is mapped to a vector of length 2. The first component of the vector is the distance to the error line $|error\_line - current\_line|$ (in out example, the error line is line 3, highlighted in red). The second component of the vector is the coded line timestamp. In our data, time is measured in milliseconds, therefore we use $\log{(report\_timestamp - line\_timestamp)}$ to account for the order of magnitude.
    \item The sequence of such vectors is processed using the RNN with attention as described in \Cref{sec:vector-representation}. 
\end{itemize}

The obtained annotation embedding can be used as an alternative to manual features extracted from annotations. 

\secpart{Similarity of Vector Representations}
\label{sec:vector-similarity}
After obtaining the embeddings of the bug and the developer, we feed them into a comparison module. Here, we have applied the approach from the work of Severyn et al.~\cite{Severyn2015LearningTR}, proposing to form the following vector:
\begin{align}
    \mathbf{x}_{join} = [ \mathbf{x}^T_q; x_{sim}; \mathbf{x}^T_d; \mathbf{x}^T_{feat} ],
\end{align}
where $\mathbf{x}_q$, $\mathbf{x}_d$, $\mathbf{x}_{feat}$ stand for the bug embedding, the developer embedding, and additional stack trace features described in Sections \ref{sec:manual-features} and \ref{sec:neural-features}. A scalar value $x_{sim}$ is obtained from $\mathbf{x}^T_q \mathbf{M} \mathbf{x}_d$ with a trainable matrix $\mathbf{M}$, which captures syntactic and semantic aspects between the queries and documents. 

After that, a feed-forward neural network with one hidden layer and ReLU activation function is applied, and the score is obtained which is used to rank developers. 
\section{Evaluation}\label{sec:evaluation}

We evaluated our approach on stack traces collected from the internal system of JetBrains, a large software company. We aim to answer the following research questions:

\textbf{RQ1:} How do ranking models work in comparison with classifying models?

\textbf{RQ2:} Do frame-based features built from VCS annotations improve the model quality? Which of them affect the performance more, the manual ones or the features learned by the neural model automatically?

\textbf{RQ3:} How does adding stack-based features to frame-based features affect the model?

\secpart{Dataset Collection}

To collect data for the evaluation, we used the crash report processing system that handles reports from various JetBrains products. When a crash occurs in the user's product (\textit{i.e.}, an IDE), an anonymous crash report is formed. If the user agreed to send such reports to the company, then it gets sent and is stored in the processing system. Since we are not able to publish internal company data, we have collected two datasets: from the company's private and public code repositories. Our datasets were created from stack traces that are automatically created after every crash of a product. The public dataset is a subset of the private dataset that contains stack traces that relate to public repositories. The public dataset is published for further research and can be found in the DapStep repository: \url{https://github.com/Sushentsev/DapStep}.

The larger, private dataset contains a total of 11,139 bug reports from the crash system from October 2018 to April 2021. These bug stack traces include files from three JVM languages: Java, Kotlin, and Scala. The proposed solution is language-agnostic, files in different languages are processed in the same way. The developer who fixed the bug in the system will be referred to as the \textit{target developer}. For each error from the dataset, the target developer is known. As mentioned earlier, we use annotations to improve the quality of our model. Annotations can be obtained from the Git version control system using the file name and the file commit hash.

The private dataset contains annotations for all files that are present in the stack trace, with the total number of annotations being 99,591. However, not all annotations are present in public repositories, only 32,908 of them. The public dataset contains stack traces, in which at least 75\% of the annotations are present publicly. This results in 3,361 different stack traces. Thus, a public dataset consists of a subset of reports from a private dataset, for which a sufficient number of annotations are available. We believe that this dataset can be useful for further research in the field and can facilitate the development of models, which work with the systems that process the reports in the form of stack traces.

\secpart{Baseline Implementations}
To compare our stack-trace-based approach with approaches that use reports description, we implement several baseline models. It is important to note that we are comparing models from the point of view of stack trace processing, because we have no textual descriptions of bugs. We apply preprocessing (\Cref{sec:preprocessing}) that converts a stack trace into a set of text tokens that can be processed as text data. As baseline models, we chose Logistic Regression and Random Forest. In addition, we have implemented a heuristic solution, which is based on counting the number of edited files by each developer. Let us describe these baselines in more detail.

Logistic regression~\cite{Hosmer1989AppliedLR} is one of the simplest and most popular machine learning models that demonstrated its capabilities in the bug triage problem~\cite{sarkar2019improving}. Logistic regression performs a linear transformation on a vector of features; to obtain the distribution of probabilities by class, the sigmoid function is used. 
In addition to logistic regression, we used Random Forest~\cite{Breiman2004RandomF} as a baseline model. Unlike Logistic Regression, Random Forests are capable of constructing a non-linear decision boundary. Thus, Random Forest is able to capture more complex data dependencies. 
We used \textit{SGDClassifier} and \textit{
RandomForestClassifier} from the \textit{scikit-learn} package as the implementations of the models.
To apply classification algorithms, each stack trace must be represented with a feature vector. One of the most popular approaches that works well in practice is the TF-IDF approach~\cite{Ramos2003UsingTT}. 

We also propose a baseline model based on a simple heuristic. For each frame of the stack trace, we know exactly which line in the file caused the bug. From the VCS annotation, we can find out which developer edited the given line last. Thus, for each developer, we count the number of edited lines that led to an error. The developer who edited the most error lines should be assigned to fix the bug. Additionally, we use the following ideas to improve the quality of this solution. Firstly, the frames at the top of the stack are usually more explanatory, therefore we can consider not all frames in the trace stack, but only Top-20 frames. Secondly, we consider each line edit with a weight that depends on the edit time: the later the line edit happened, the higher the weight is. As a weight function, we used $f(x) = e^{-x}$, where $x$ stands for the time elapsed from editing a line until a bug occurred in the system.

\secpart{Model Parameters}

Since we collected two different datasets from public and private repositories, for each dataset, the parameters of the models were selected independently. The model parameters are selected according to the results on the validation datasets.

\begin{table}[t]
    \centering
    \caption{Parameters used for different models.}
    \label{table:model_parameters}
    \begin{tabular}{lcc}
        \toprule
        \textbf{Parameter} & \textbf{Public dataset} & \textbf{Private dataset} \\
        \midrule
        \multicolumn{3}{c}{\textbf{Logistic Regression}} \\\midrule
        \textbf{Loss} & log & log \\
        \textbf{Regularization coefficient} & 1e-5 & 1e-5 \\\midrule
        
        \multicolumn{3}{c}{\textbf{Random Forest}} \\\midrule
        \textbf{Number of estimators} & 100 & 100 \\
        \textbf{Maximum depth} & $\infty$ & $ \infty$\\
        \textbf{Minimum samples in a leaf} & 1 & 1 \\\midrule
        
        \multicolumn{3}{c}{\textbf{CNN}} \\\midrule
        \textbf{Number of convolutional filters} & 32 & 64 \\
        \textbf{Size of trainable embeddings} & 50 & 70\\\midrule
        
        \multicolumn{3}{c}{\textbf{RNN}} \\\midrule
        \textbf{Hidden size} & 70 & 100 \\
        \textbf{Size of trainable embeddings} & 50 & 70\\
        \bottomrule
    \end{tabular}
\end{table}

The detailed information about the parameters can be found in~\Cref{table:model_parameters}.
In the proposed neural network models, the dropout~\cite{Srivastava2014DropoutAS} and weight decay~\cite{Loshchilov2019DecoupledWD} are applied to prevent overfitting. We used the Adam optimizer~\cite{Kingma2015AdamAM} with a learning rate of $1\mathrm{e}{-3}$ and a weight decay of $1\mathrm{e}{-3}$, dropout rate was $0.2$. The classifying models were trained for 25 epochs, and the ranking models were trained for 10 epochs because our experiments have shown that a larger number of epochs leads to the model overfitting.

\secpart{Loss Function}

Since we considered bug triage as a ranking problem, it is necessary to prepare labels for the ranking problem: the target developer must be the first in the list of the ranked developers. For our problem statement, a pairwise approach to RankNet~\cite{Burges2005LearningTR} loss is often used.

The RankNet algorithm assumes that the training data consists of pairs of documents $(d_1, d_2)$ together with a target probability $\bar{P}$ of $d_1$ being ranked higher than $d_2$. For each query, there is only one relevant document (target developer), all other documents (developers) are considered irrelevant. 

As a result, the final loss function with simplification for several pairs $(d_i, d_j)$ and query $q$ has the following form: 
\begin{align}\label{eq:loss}
    L = \sum_{d_i \prec d_j} \log{\left (1 + e^{-(f(q, d_i) - f(q, d_j))} \right)}
\end{align}

To evaluate our approach, we take a random query (bug stack trace) $q$ and a set of documents (developer vector representations) $\{d_1, \ldots, d_n \}$, 
and make a gradient descent step according to \eqref{eq:loss}.
Furthermore, we use the following heuristic observation: if the developer stack trace is empty, then they did not edit any file from the bug stack trace. It is unlikely that this developer will fix the current bug, therefore, we exclude such a developer from the list of possible assignees. It is also essential that the calculation of the loss function requires the score of the target developer. However, the target developer representation in the stack trace form may be empty, therefore, in such cases we remove such reports from the training data.

\secpart{Performance Metrics}

\begin{table*}
    \centering
    \caption{Comparison of the models on the public and private datasets.}
    \begin{tabular}{clcccccccc} 
        \toprule
        \multicolumn{1}{c}{\multirow{2}{*}{\textbf{№}}} & \multicolumn{1}{c}{\multirow{2}{*}{\textbf{Model}}} & \multicolumn{4}{c}{\textbf{Public dataset}} & \multicolumn{4}{c}{\textbf{Private dataset}} \\ 
        \cmidrule(lr){3-6}\cmidrule(lr){7-10}
        && \textbf{Acc@1} 
        & \textbf{Acc@5} & \textbf{Acc@10} & \textbf{MRR} & \textbf{Acc@1} 
        & \textbf{Acc@5} & \textbf{Acc@10} & \textbf{MRR} \\ 
        \midrule
        1 &Heuristics & 0.26 
        & 0.50	& 0.52 & 0.36 & 0.41 
        & 0.73 & 0.80 & 0.54 \\ 
        
        2 & Logistic Regression & 0.19 
        & 0.35 & 0.45 & 0.27 & 0.43 
        & 0.56 & 0.62 & 0.50 \\ 
        
        3 & Random Forest & 0.16 
        & 0.33 & 0.40 & 0.25 & 0.42 
        & 0.57 & 0.64 & 0.50 \\ 
        
        4 & CNN classification & 0.14 
        & 0.29 & 0.38 & 0.22	 & 0.42 
        & 0.55 & 0.60 & 0.48 \\ 
        
        5 & RNN classification & 0.14 
        & 0.27	& 0.34 & 0.21	& 0.42  
        & 0.54 & 0.60 & 0.48 \\ 
        
        6 & CNN ranking (without VCS) & 0.13
        & 0.37 &	0.47 & 0.25 & 0.35 
        & 0.60 & 0.72 & 0.48 \\ 
        
        7 & RNN ranking (without VCS) & 0.21 
        & 0.37 &	0.50 & 0.30 & 0.46 
        & 0.69 & 0.76 & 0.57 \\ 
        \midrule
        8 &CNN ranking (VCS: manual frame-based) & 0.28	
        & 0.48 & 0.54 & 0.38 & 0.53 
        & 0.79 & 	0.84 & 0.65 \\ 
        
        9 & CNN ranking (VCS: neural frame-based) & 0.29 
        & 0.48 & 0.54 & 0.39 & 0.54
        & 0.80 & 0.84 & 0.66 \\ 
        
        10 & RNN ranking (VCS: manual frame-based) & \textbf{0.35} 
        & \textbf{0.52}	 & \textbf{0.60}	& \textbf{0.44} & 0.58 
        & 0.82 & 0.86 & 0.68 \\ 
        
        11 & RNN ranking (VCS: neural frame-based) & 0.27 
        & 0.46 & 0.56 & 0.37 & 0.54 
        & 0.79 & 0.83 & 0.65 \\ 
        \midrule
        12 & CNN ranking (VCS: manual frame-based \& stack-based) & 0.31 
        & 0.49 &	0.56 & 0.40 & 0.57 
        & 0.82 & \textbf{0.87} & 0.68 \\ 
        
        13 & RNN ranking (VCS: manual frame-based \& stack-based) & 0.34 
        & \textbf{0.52} & 0.56 & 0.43 & \textbf{0.60} 
        & \textbf{0.83} & \textbf{0.87} & \textbf{0.70} \\ 
        \bottomrule
    \end{tabular}
    \label{table:results}
\end{table*}

To answer the research questions, we compared the proposed ranking model with the classification models adapted for stack traces. The most common quality metric for classification problems is Accuracy at K. Accuracy at K corresponds to the number of relevant results among the first $K$ positions. However, this metric does not take into account the position of the relevant document, therefore, we used different values of $k$ from the $\{ 1, 5, 10 \}$ set. More formally, the Accuracy at K metric is defined as follows: 
\begin{align}
    acc@k = \frac{1}{|D|} \sum_{(d, q) \in D} \mathbb{I} \left(d \in \{d_{q_i} \}_{i = 1}^{k} \right),
\end{align}
where $\mathbb{I}$ stands for the indicator function and $\{d_{q_i}\}$ is the ranked list of documents for query $q$.

In the ranking problem, we use mean reciprocal rank (MRR) for evaluation, which corresponds to the harmonic mean of the relevant documents' ranks. It should be noted that only the rank of the first relevant document is considered in MRR. However, it is suitable for our task, since there is always one relevant document for each query. MRR can be calculated using the following formula: 
\begin{align}
    MRR = \frac{1}{|D|} \sum_{(d, q) \in D} \frac{1}{rank_d^q},
\end{align}
where $rank_d^q$ refers to the rank position of the target document $d$ for the query $q$.

\secpart{Experiment Methodology}

To evaluate our models, we divided both datasets into three sets: train, validation, and test. For the private dataset, the sizes of the train, validation, and test sets were 8139, 1500, and 1500 bug stack traces, respectively. For the public dataset, the split was 2461, 450, and 450, respectively. This corresponds to the validation and test sets being about 15\% of the sizes of the entire datasets, which is a common practice. This partitioning helps to prevent overfitting of the model. Since the data has a time component, the dataset is divided by time in order to avoid data leakage.

Our methodology for the experiment with the classification models is as follows: we select hyperparameters using the validation datasets, then fit the model on the training and validation datasets, and, finally, evaluate the quality of the models on the test datasets. If the developer is found only in the test dataset, then we cannot correctly classify the bug, since the model was not trained for this class. In this case, we consider that the bug was assigned incorrectly.

For the ranking problem, the model was evaluated as follows. During the training, a random stack trace is taken from the training dataset. Then, for each developer, their stack trace representation is built. If the target developer has an empty stack trace representation, then this means that the developer did not fix frames from this stack trace. In this case, we exclude this stack trace from the training dataset. When evaluating, the model considers only those developers whose stack trace representation is not empty. If the developer's stack trace representation is empty, then his score is equal to the minimum score predicted by the model.

To test the statistical significance of our results, we use bootstrap~\cite{Efron1979BootstrapMA} to construct the confidence intervals. When comparing two models, we form 100 bootstrapping resamplings with the same size as the test dataset. Next, a 95\% confidence interval for the difference of the metric scores is calculated. If zero falls into the constructed interval, then there are no statistically significant differences between the models, otherwise, we say that there is statistical significance. 

All the experiments were run on a server with the following technical characteristics: 8-core Intel Xeon CPU @2.3 GHz, NVidia K80 GPU, and 60 GB of RAM.

\secpart{Results}

The experimental results of running various models on both datasets are presented in~\Cref{table:results}. Resulting confidence intervals for all the experiments can be found in the online appendix: \url{https://doi.org/10.5281/zenodo.5596294}.

First of all, it can be seen that the results are different for the public and the private datasets. We assume that this happened for three reasons. Firstly, the public dataset is several times smaller than the private dataset, which can affect the approaches that rely on a lot of data. Secondly, not all annotations are available for the public dataset, with the missing annotations likely containing some important information. Thirdly, we found that the test set from the public dataset contains more target developers who have not edited files from stack traces than the private dataset. Thus, their stack trace representation will be empty, and the result of the model will be incorrect on these reports.

\subsubsection{Research Question 1} 
To answer RQ1, let us evaluate and compare the quality of the classifying and ranking models. Our results show that classifying models based on classical machine learning algorithms perform as well as classifying algorithms based on RNNs or CNNs (\Cref{table:results}, lines 2--5). We believe that this can be explained by the fact that neural networks are most likely to extract features similar to TF-IDF features, so the results are similar. 

The RNN ranking model performs better than the others (\Cref{table:results}, line 7, 0.21 Acc@1 on the Public dataset, 0.46 Acc@1 on the Private dataset), but the differences would be more significant if there were many developers in the test dataset that were not in the training dataset. We found that for the public and private datasets, there were 27 bug reports in the test data with developers that were not presented in train data. Thus, this represents only 6\% and 1.8\% of the total size of the test data, and the advantage of the ranking approach is not very noticeable. On the other hand, for projects that have a larger variety of developers (for example, some large open-source projects) the ranking approach can be expected to significantly improve the quality of the models.

Another interesting observation is that the heuristic-based baseline shows the quality comparable to the ML-based approaches. Such high performance of the heuristic solution can be explained by the fact that the data was collected from industrial projects of a large software company with well-functioning bug fixing pipelines, and the proposed heuristic might be a good fit for such a pipeline. In open-source projects, error processing workflows might be different, and as a result, such a heuristic solution might work worse. However, this observation suggests that sometimes a simple heuristic might work better than complex statistical models that are not interpretable and need a lot of sophisticated data to train on.

\observation{On our datasets, the ranking models perform slightly better, but the difference can be more significant in other settings, future research is required.}

\subsubsection{Research Question 2} 
Next, let us address RQ2 and investigate the significance of manual and neural frame-based features built from the VCS annotations. We trained a ranking model with only manual frame-based features and another model with only neural frame-based features. It can be seen (\Cref{table:results}, lines 8--11 compared to lines 6--7) that frame-based features increase the model quality, but the impact of the neural features is not as significant as in the case of manually extracted features in the RNN model (0.27 and 0.35 Acc@1 on the public dataset, 0.54 and 0.58 Acc@1 on the private dataset, respectively). However, in the case of the CNN-based approaches, manual and neural features show similar results. CNN captures the entire stack trace, rather than processing it in a sequential form like the RNN does. Therefore, feature normalization in the case of CNN may be necessary, since a lot of raw values are harmful. The difference between manual frame-based features and neural frame-based features turned out to be statistically significant for RNN and not significant for CNN on both datasets.

An important disadvantage of the neural network annotation processing is the slow model training (one epoch takes 3-4 times longer compared to the manual features): each pass requires hundreds of annotations to be processed, each of them can contain thousands of lines, and since we use RNN, it takes a significant amount of time. On the other hand, the DL approach learns annotation embeddings automatically, and these embeddings could be useful in other related tasks (bug localization, bug report deduplication, etc.). This seems like a promising direction for future work.

\observation{Adding frame-based features from the VCS annotations improves the quality of models. Manual features worked better for the RNN models, while in CNN models, the difference between manual and learned features is insignificant. Learning features takes noticeably more time, but leads to obtaining embeddings of annotations, which might be useful for other tasks.}

\subsubsection{Research Question 3} 

Finally, to answer RQ3, we trained models with both the frame-based and stack-based features from the VCS annotations. Since manual frame-based features demonstrated better results than neural features, we only used manual features. First, we can notice that the RNN-based model outperforms the CNN-based one by 3 percentage points according to Acc@1 (\Cref{table:results}, lines 12--13), however, in the case of the public dataset, this difference is not statistically significant. 
Better performance of RNNs compared to CNNs may be attributed to the CNN training pipeline: to reduce the training time, stack traces are processed in batches. At the same time, CNN is not designed to process sequences of different lengths in batches, therefore, padding is necessary.  Moreover, the length of the sequences must not be shorter than the size of the convolution kernel, that is, 5. It is possible that padding in the training data leads to worse results. 

Secondly, we can see that adding stack-based features has a positive statistically significant effect on the model performance (\Cref{table:results}, lines 8, 10, 12--13). 
We believe that frame-based features are not taken into account in the best way in CNN models, therefore, stack-based features add new information to the model. However, in the case of RNN models, stack-based features do not lead to such improvements. Perhaps, better feature engineering could help us overcome this issue, future research is required.

\observation{Combining stack-based and frame-based has a positive effect on the CNN-based appoaches, but for the RNN-based models the effect is not significant.}

In the end, the RNN ranking model with frame-based and stack-based manual features obtained from the VCS annotations turned out to be the best performing model for bug assignee prediction based on the stack traces data. 
It outperforms all the other models on the private dataset (\Cref{table:results}, line 13, 0.60 Acc@1 and 0.70 MRR) and achieves a significant boost in all metrics (17--18 percentage points) compared to the classical machine learning approaches. We release this model as \textit{DapStep} and plan to conduct more thorough experiments on it in the future work.

Thus, the results of our experiments demonstrate that reformulating bug triage as a ranking problem is appropriate. Moreover, adding features from VCS annotations to the model has a positive effect on its performance, and the RNN-based models work slightly better in this setting than the CNN-based ones. From the practical standpoint, the RNN ranking model with all the features can be trained on the data of any specific project or company and be employed there. As for the research implications, the results show that more research is needed to improve the state-of-the-art solutions to the bug triage problem, employing more information about the stack traces. We hope that our results of using VCS annotations as the sources of features and the provided dataset can assist in conducting such research.
\section{Threats to validity}\label{sec:threats-to-validity}
Our study suffers from the following threats to validity.

\textbf{Subject selection bias.} The performance of the model depends on the data. Since stack traces for the bug triage task are being used for the first time, there is no dataset available for this task. We collected a dataset from the products of a large software company and evaluated the proposed approach on them. However, applying the model to other dataset may lead to different results. For instance, workflows in open-source projects could be more volatile and unstable. The results for such datasets can be noticeably lower.

\textbf{Limited scope of application.} Our solution is applicable for software systems that report stack traces when a bug happens, which might be not be typical for some projects and companies. However, we believe this practice to be common enough for our approach to be helpful in practice. 
Secondly, deep learning models are over-parameterized. A modern neural network contains thousands or millions of parameters. A sufficient amount of data is required to train a neural network. We use 11,139 different stack traces in our private dataset and regularization techniques to prevent overfitting. However, in cases when this amount of data is not available, the results may differ. We hope our research will encourage other researchers and practitioners to invest time and effort into collecting a larger dataset of such kind.

\textbf{Programming language bias.} Our datasets consist of stack traces that were obtained from the JVM languages. Therefore, the results of our models for other languages may differ. Firstly, stack trace characteristics change from one language to another. The performance of the model depends on the average length of the stack trace, as well as the variety of methods and files used. Secondly, an essential component of our approach is the use of features from annotations. The characteristics of these files also strongly affect the model performance. The distribution of developers for each file can vary between teams, companies, and maybe even programming languages. Future research is needed to assess how much all of this affects the resulting model. 

While these threats to validity are important to note, we believe that they do not invalidate the overal results of our study and its practical usefulness.
\section{Conclusion and future work}\label{sec:conclusion}

In this paper, we explore the applicability of using stack traces for solving the bug triage problem.

Firstly, we suggest an approach for handling error reports that do not have text descriptions, but only a stack trace for the given error. We transform each stack trace into a set of text tokens, which are processed as sequences. As a result, existing solutions can be applied to such data as well.

Secondly, we collected two datasets---the public one and the private one---from the data of JetBrains. The public dataset is a subset of the private dataset that only contains stack frames that relate to public repositories, with a total of 3,361 stack traces. To facilitate further research in this area, the source code of all the models, as well as the public dataset, are available online at \url{https://github.com/Sushentsev/DapStep}.

Thirdly, we propose a ranking neural network model that outperforms classifying models by 15-20 percentage points of the Acc@1 metric on the public dataset, and 17-18 percentage points on the private dataset. The significant advantage of this model is the independence from the fixed set of classes (the list of developers working on a given project). Finally, we suggest to use an additional source of information (VSC annotations), which significantly improves the performance of the models. We propose two ways features could be built from such annotations. First of all, features can be extracted manually from annotations --- this approach shows better results, but requires effort and domain knowledge. On the other hand, it is possible to use an additional neural network to learn the annotation-based features. This approach requires to train an additional neural network, so it takes more time compared to the manual approach, however, this way we obtain explicit embeddings of annotations, which might be employed in other related research tasks.

We hope that our work will be of use for researchers and practitioners, especially in the tasks that rely on stack traces.

\balance
\bibliographystyle{IEEEtran}
\bibliography{IEEEabrv,paper}
\end{document}